\pdfoutput=1

\documentclass[11pt]{article}

\usepackage[preprint]{acl}

\usepackage{times}
\usepackage{latexsym}

\usepackage[T1]{fontenc}

\usepackage[utf8]{inputenc}

\usepackage{microtype}

\usepackage{inconsolata}

\usepackage{graphicx}

\usepackage{amsmath}
\usepackage[capitalise, noabbrev]{cleveref}
\usepackage{bm}
\usepackage{enumitem}
\usepackage{xspace}
\usepackage{multirow}
\usepackage{graphicx}
\usepackage{caption}
\usepackage{float}
\usepackage{subcaption}
\usepackage[ruled,linesnumbered]{algorithm2e}
\usepackage{algpseudocode}
\usepackage{makecell}
\usepackage{hyperref}
\usepackage{url}
\usepackage{dsfont}
\usepackage{colortbl}
\usepackage{tikz}
\usepackage{setspace}
\usepackage{natbib}
\usepackage{booktabs}

\newtheorem{prbl}{\bf Problem}

\newcommand{\ie}{\textit{i.e.,}\xspace}

\newcommand{\eg}{\textit{e.g.,}\xspace}

\newcommand{\model}{\textsc{SCoRe}\xspace}
\newcommand{\ml}{ML-1M\xspace}
\newcommand{\games}{Games\xspace}

\newcommand{\rev}[1]{\textcolor{black}{#1}}
\newcommand{\emnlp}[1]{\textcolor{black}{#1}}

\newcommand{\rep}[1]{\bm{#1}}

\newcommand{\mybox}[2]{
    \begin{figure}[H]
        \centering
    \begin{tikzpicture}
        \node[anchor=text,text width=0.93\columnwidth, draw, rounded corners, line width=0.8pt, fill=gray!10, inner sep=2mm, align=justify] (big) {
        	\begin{spacing}{0.8}
        	\vspace{0.3ex}
         {\small #2}
         	\vspace{-2.5ex}
        	\end{spacing}
        };
        \node[draw, rounded corners, line width=1pt, fill=gray!40, anchor=west, xshift=3mm] (small) at (big.north west) {{\small #1}};
    \end{tikzpicture}
    \vspace{-2mm}
    \end{figure}
}

\title{LLMs as Better Recommenders with Natural Language Collaborative Signals: A Self-Assessing Retrieval Approach}

\author{
 \textbf{Haoran Xin\textsuperscript{1}}\quad
 \textbf{Ying Sun\textsuperscript{1}}\quad
 \textbf{Chao Wang\textsuperscript{2}}\quad
 \textbf{Weijia Zhang\textsuperscript{1}}\quad
 \textbf{Hui Xiong\textsuperscript{1}}\quad
\\
 \textsuperscript{1}Thrust of Artificial Intelligence, The Hong Kong University of Science and\\
 Technology (Guangzhou)
 \textsuperscript{2}School of Artificial Intelligence and Data Science,\\
 University of Science and
 Technology of China
\\
\texttt{\{hxin883, wzhang411\}@connect.hkust-gz.edu.cn, yings@hkust-gz.edu.cn} \\
  \texttt{wangchaoai@ustc.edu.cn,}
  \texttt{xionghui@ust.hk}
}

\begin{document}
\maketitle

\begin{abstract}
\emnlp{
Incorporating collaborative information~(CI) effectively is crucial for leveraging LLMs in recommendation tasks.
Existing approaches often encode CI using soft tokens or abstract identifiers, which introduces a semantic misalignment with the LLM's natural language pretraining and hampers knowledge integration.
    To address this, we propose expressing CI directly in natural language to better align with LLMs' semantic space.
    We achieve this by retrieving a curated set of the most relevant user behaviors in natural language form.
    However, identifying informative CI is challenging due to the complexity of similarity and utility assessment.
    To tackle this, we introduce a \underline{S}elf-assessing \underline{\textsc{Co}}llaborative \underline{\textsc{Re}}trieval framework (\model)
    following the retrieve-rerank paradigm.
    First, a Collaborative Retriever~(CAR) is developed to consider both collaborative patterns and semantic similarity.
    Then, a Self-assessing Reranker~(SARE) leverages LLMs’ own reasoning to assess and prioritize retrieved behaviors.
Finally, the selected behaviors are prepended to the LLM prompt as natural-language CI to guide recommendation.
Extensive experiments on two public datasets validate the effectiveness of \model in improving LLM-based recommendation.}

\end{abstract}

\section{Introduction}

As large language models (LLMs) have shown impressive world knowledge and reasoning capabilities, researchers have sought to extend the potential of LLMs to recommender systems (RSs), leveraging their ability to understand user behavior semantics~\cite{wu2024survey,dai2023uncovering}.
However, despite their proficiency in general knowledge-intensive tasks, LLMs face inherent limitations in recommendation scenarios 
\rev{due to their pre-trained knowledge base lacking crucial collaborative information (CI) embedded in user-item interactions~\cite{zhang2023collm}.}

\begin{figure}[t]
    \centering
    \includegraphics[width=0.7\columnwidth]{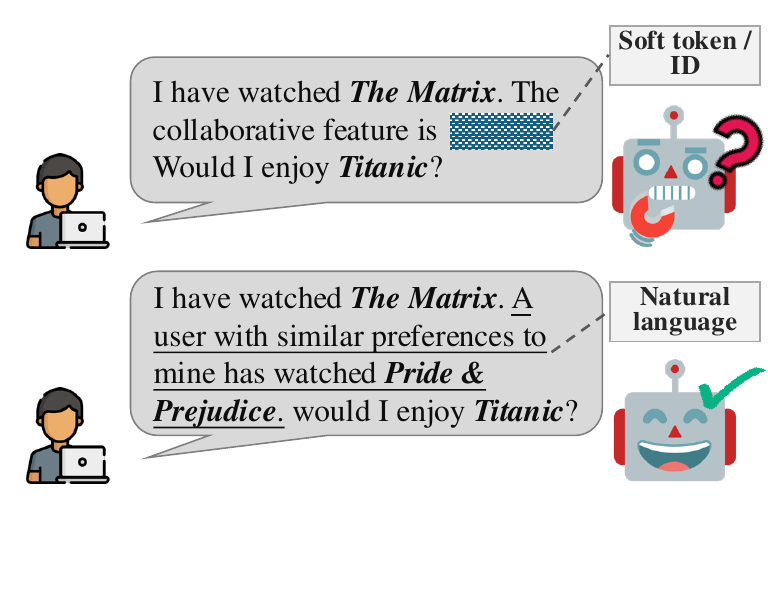}
    \vspace{-8mm}
    \caption{The use of soft tokens or less meaningful identifiers as CI introduces a semantic gap that impairs LLM understanding. In contrast, LLMs could more effectively comprehend CI conveyed in natural language.}
    \vspace{-6mm}
    \label{fig:intro}
\end{figure}

To enhance LLMs' recommendation capabilities, recent works~\cite{liao2024llara,zhang2023collm,zheng2024adapting} have explored methods to incorporate CI into LLMs.
Given that conventional recommendation models (CRMs) effectively capture CI using latent factor models such as matrix factorization~\cite{rendle2012bpr}, multiple works~\cite{liao2024llara,zhang2023collm,kim2024large} aim to adapt the collaborative knowledge learned from CRMs into LLMs by projecting latent embeddings into LLM tokens.
These learned tokens are then inserted into input prompts to represent CI for final recommendation generation.
\rev{Recent advances~\cite{jia2024bridging,zhang2024text} further enhance this approach by representing users and items through explicit identifiers or binary encodings within the token space.}

\emnlp{However, representing CI using soft tokens or abstract identifiers introduces a fundamental modality gap with LLMs, which are pre-trained on \textit{natural language corpora}. This semantic misalignment limits the models’ ability to interpret and leverage CI in recommendation tasks.}
\emnlp{
As a result, expressing CI directly in natural language emerges as a promising direction, as it better aligns with LLMs’ pretraining and enhances their capacity to understand and utilize collaborative signals.
}


\rev{As retrieval-augmented generation (RAG)~\cite{jia2024bridging,gao2023retrieval,zhang2023retrieve} enabling non-parametric integration of external information into LLMs advances, researchers~\cite{lin2024rella,wu2024coral} have explored retrieval-based CI incorporation to mitigate semantic divergence.
Notably, \citet{wu2024coral} propose retrieving user-item pairs as supplementary CI.
However, their approach presents retrieved pairs as identifiers, still potentially impeding LLMs' comprehension.}
\rev{This limitation motivates our novel research question: \textit{Can we effectively incorporate CI by retrieving similar user behaviors in natural language form?}
This approach could potentially maximize LLMs' understanding by aligning their pretraining on natural language. 
\emnlp{
Since item metadata (\eg textual descriptions) is typically easy to access, generating natural-language representations of user behaviors is feasible.
The core challenge then shifts to how to retrieve behaviors that are the most relevant to the target user.%
}
However, this direction presents several fundamental challenges:
First, \textbf{measuring behavioral similarity in LLM-based RSs is inherently complex.}}
Unlike CRMs that primarily focus on behavioral pattern similarity, LLMs necessitate consideration of both interaction patterns and the semantic context embedded within these interactions, making the similarity assessment more complicated and multifaceted.
\emnlp{
Second, \textbf{identifying which retrieved behaviors offer beneficial collaborative signals for the recommendation task remains an open problem.}
Although many similar user behaviors can be retrieved, not all of them contribute meaningfully to improving recommendations.
The utility role of historical user behaviors in enhancing LLM-oriented recommendation remains insufficiently understood.
}

\emnlp{To address these challenges, we propose a \underline{S}elf-assessing \underline{\textsc{Co}}llaborative \underline{\textsc{Re}}trieval framework (\model) to retrieve similar user behaviors, and then represent them in natural language to enhance LLM-based recommendation. 
\model adopts a two-stage retrieve-rerank paradigm, consisting of a Collaborative Retriever (CAR) and a Self-Assessing Reranker (SARE), through a two-stage lightweight fine-tuning.
In the first stage, CAR is trained using collaborative rankings derived from a pretrained CRM, leveraging an off-the-shelf retriever.
This enables CAR to capture user similarity by jointly modeling collaborative interaction patterns and semantic context.
In the second stage, we harness the LLM’s own reasoning to assess the characteristics of useful collaborative signals.
The generated assessments guide the fine-tuning of SARE, a general-purpose reranker adapted to prioritize behaviors likely to enhance recommendation quality.
During inference, CAR retrieves a candidate set of similar user behaviors, which SARE reranks based on their potential utility.
The top-ranked behaviors are then converted into natural language and prepended to the LLM prompt as supplementary CI, effectively improving the model's recommendation capability.
The main contributions of this work are highlighted as follows:}
\begin{itemize}
	\item \emnlp{We propose representing CI in natural language form to bridge the semantic gap between CI and LLMs, thereby enhancing the LLM-based recommendation performance.}
	\item \emnlp{We introduce \model, a retrieval framework that identifies informative user behaviors as CI by integrating joint similarity and leveraging LLM-driven self-assessment.}
	\item \emnlp{We conduct extensive experiments on two datasets to validate the effectiveness of our proposed framework.}
\end{itemize}

\section{Problem Formulation}
We begin by formally defining the problem addressed in this work.
Let $\mathcal{U}$ and $\mathcal{V}$ denote the user and item sets, respectively. The user-item interaction matrix is represented as $\mathcal{Y} \in \{0,1\} ^ {|\mathcal{U}| \times |\mathcal{V}|}$, where each entry $y_{uv} \in \{0,1\}$ indicates whether user $u \in \mathcal{U}$ has interacted with item $v \in \mathcal{V}$. Specifically, $y_{uv}=1$ denotes a positive interaction, and $y_{uv}=0$ otherwise.
Based on the interaction matrix $\mathcal{Y}$, we define the sequence of interacted items $H(u)$ for a given user $u \in \mathcal{U}$ as:
\begin{equation}
\begin{aligned}
  H(u) &= \left[ v_1, \ldots, v_{|H(u)|} \right], \\
  \forall v \in H&(u), v \in \mathcal{V} \wedge y_{uv} = 1.
\end{aligned}
\end{equation}
The complete interaction corpus, denoted as $\mathcal{H}$, comprises the set of all user interaction sequences $H(u)$.
\begin{figure*}[t]
  \centering
  \includegraphics[width=0.93\textwidth]{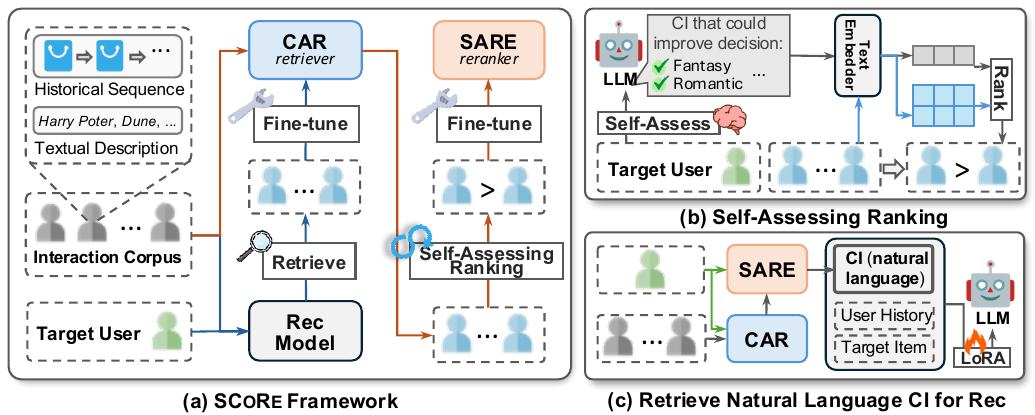}
  \vspace{-2mm}
  \caption{
	(a) Overview of the \model framework. A two-stage fine-tuning paradigm is used to develop the collaborative retriever (CAR) and the self-assessing reranker (SARE).
	(b) Illustration of the self-assessing ranking process. The LLM evaluates the characteristics of beneficial CI for recommendation, which guides the reranking of retrieved similar users.
	(c) User behaviors are retrieved and reranked by CAR and SARE, then prepended to the prompt in natural language to enhance LLM-based recommendations.
  }
  \label{fig:framework}
  \vspace{-7mm}
\end{figure*}
In this work, we first retrieve relevant similar user behaviors from the interaction corpus as supplementary CI, which is then utilized in natural language form by an LLM for recommendation generation. The problem is formally defined as:
%
\begin{prbl}
\textit{
Given the user set $\mathcal{U}$, the item set $\mathcal{V}$, the interaction matrix $\mathcal{Y}$ and the interaction corpus $\mathcal{H}$, design a retrieval strategy $\mathcal{R}(\cdot)$ to retrieve similar users and their behaviors in $\mathcal{H}$ that serve as supplementary CI.
Then, for a target user $u_t$ with interactions $H(u_t)$, the retrieved users $\mathcal{S}(u_t) \subset \mathcal{U}$ and their behaviors are used in natural language form to augment an LLM for predicting the probability $\hat{y}_{u_tv_t}$ (denoted as $\hat{y}_t$ for brevity) of the user $u_t$ liking a target item $v_t$, \ie}
\begin{equation}
  \small
\begin{aligned}
\mathcal{S}(u_t) = \mathcal{R}(&u_t, v_t),~ \mathcal{C}(u_t) = \{d(H(u_i))|u_i \in \mathcal{S}(u_t)\}, \\
	&\texttt{LLM}(d(u_t),d(v_t),\mathcal{C}(u_t)) \mapsto \hat{y}_{t},
\end{aligned}
\end{equation}
\textit{where $d(\cdot)$ denotes the operation of converting the input into natural language using item metadata.}
\end{prbl}

\section{Methodology}
In this section, we present the proposed \model for retrieving CI in natural language format for LLM-based recommendation. We first introduce the overall framework and then describe the details.

\subsection{Framework Overview}
\cref{fig:framework}(a) presents an overview of our proposed framework, \model, which constructs CAR and SARE to retrieve and refine similar user behaviors as CI.
First, \textit{CollAborative Retriever (CAR)} is developed by fine-tuning a semantic-rich retriever using collaborative rankings derived from a pretrained CRM. This enables the retrieval of user behaviors that are similar both collaboratively and semantically~(\cref{sec:retriever}).
\emnlp{Next, as illustrated in Figure 2(b), we introduce a self-assessing ranking strategy that prompts the LLM to assess the characteristics of beneficial CI for recommendation.
These assessment outputs are then used to rank retrieved behaviors and fine-tune \textit{Self-Assessing REranker~(SARE)} accordingly~(\cref{sec:reranker}).
Finally, during inference, the optimized CAR and SARE sequentially retrieve and rerank relevant user behaviors, which are then expressed in natural language and prepended to the prompt as supplementary CI, enhancing the LLM’s recommendation quality~(\cref{sec:recommendation}).}

\subsection{Collaborative Retriever}
\label{sec:retriever}
When using LLMs for recommendations, the results are fundamentally driven by the models’ semantic comprehension capabilities.
Consequently, when integrating similar user behaviors as collaborative knowledge, it is crucial to account for both semantic context and behavioral patterns.
To address this, we propose a \textit{CollAborative Retriever (CAR)} that integrates both collaborative patterns and semantic information to retrieve similar user behaviors from the interaction corpus.

Given the interaction corpus $\mathcal{H}$, we first use a pretrained CRM (\eg SASRec) to generate embeddings for each user's interaction sequence $H(u)$:
\begin{equation}
	\small
	\rep{h} = \texttt{CRM}\left( H(u) \right), u \in \mathcal{U}.
\end{equation}
For a target user $u_t$, the $K_c$ most similar users $\mathcal{N}^{c}(u_t)$ are retrieved from the corpus by calculating collaborative similarity via the inner product:
\begin{equation}
	\small
\begin{aligned}
	\mathcal{N}^{c}(u_t) &= \left\{u_1, u_2, \ldots ,u_{K_c                       }\right\} \\
	&= 
	\underset{i=1,\ldots,t-1,t+1,\ldots,|\mathcal{U}|}{\texttt{Top-K}}
 \left( \rep{h}_i^{\mathrm{T}} \cdot \rep{h}_t \right) ,
\end{aligned}
\end{equation}
Next, we incorporate the textual descriptions of items to convert the interactions of the target user and the retrieved similar users into textual sentences.
To capture the semantic information embedded in these interactions, we utilize the encoder of an off-the-shelf retriever to transform these textual representations into embeddings:
\begin{equation}
	\small
\begin{aligned}
	\rep{e}^r_t = \texttt{RetrieverEnc}(d(H(u_t))),& \\
	\rep{e}^r_i = \texttt{RetrieverEnc}(d(H(u_i))), u_i \in & \mathcal{N}^c(u_t),
\end{aligned}
\end{equation}
%
To enhance the retriever with collaborative awareness, we employ a contrastive learning objective~\cite{chen2020simple} to fine-tune the retriever:
\begin{equation}
\small
\label{eq:car}
\begin{aligned}
	&\mathcal{L}_{CAR}= - \sum_{u_i \in \mathcal{N}^{c}(u_t)} \log \frac{g(\rep{e}^r_{t},{e}^{r}_i)}{g(\rep{e}^r_{t},{e}^{r}_i) + \sum_{j \in \overline{\mathcal{N}}^c{(u_t)}} g(\rep{e}^r_{t},{e}^{r}_j) },
\end{aligned}
\end{equation}
where $g(A,B) = \mathrm{exp}\left({ A \otimes B / \tau }\right)$,
$\otimes$ represents cosine similarity, $\overline{\mathcal{N}}^c$ refers to the set of all in-batch negative samples, and $\tau$ is the temperature.
\emnlp{
The resulting CAR effectively retrieves similar users by jointly capturing collaborative interaction patterns and semantic context, enabling a more accurate and comprehensive assessment of user similarity tailored to LLM-based recommendation.}

\subsection{Self-Assessing Reranker}
\label{sec:reranker}
\emnlp{
Once similar users are retrieved by CAR, it becomes essential to determine which behavioral segments are most beneficial for enhancing the LLM's recommendation performance.
This step is particularly critical given LLMs’ limited context window and their vulnerability to the lost-in-the-middle issue~\cite{liu2024lost}.
To address this, we leverage the reasoning capabilities of LLMs by prompting them to assess the characteristics of effective collaborative information.
The resulting assessments are then used to generate a relevance ranking over the retrieved user behaviors.
These rankings serve as supervision signals to fine-tune the \textit{Self-Assessing REranker (SARE)}, built on a general-purpose reranker architecture.}
Using the similar behaviors retrieved by CAR, SARE effectively prioritizes behavioral patterns that demonstrate higher potential utility for LLM recommendation and finally improves the performance.


{\textbf{LLM Self-Assessment.}}
To obtain the characteristics of beneficial CI in LLM recommendation, we present the target user-item pair $(u_t, v_t)$ to an LLM with a self-assessing prompt $P^r$:
\begin{equation}
	\small
	r_t = \texttt{LLM} \left(P^r\left(d(H(u_t)), d(v_t)\right)\right),
\end{equation}
where 
$r_t$ represents the generated assessment thoughts.
Specifically, we prompt the LLM to assume that the current information is insufficient to make a definitive recommendation and request it to identify additional supplementary characteristics related to the user’s preferences that could enhance recommendation precision.
Taking the movie recommendation as an example, the prompt $P^r$ is:

\vspace{-1mm}
\mybox{Self-Assessing Prompt $P^r$}{
The user has watched the following movies: \textit{<HistoryList>}. However, based on this list alone, it is not possible to confidently predict whether they would enjoy the movie \textit{<TargetItem>}. What other genres or characteristics related to their preferences, apart from the given history, could help in making a more informed decision?
}
\vspace{-2mm}
\noindent
An LLM assessment example is in \cref{app:example}.

{\textbf{Ranking Generation.}}
With a well-trained CAR, we retrieve $K_{e}$ similar users, the set of which is denoted as $\mathcal{N}^{e}(u_t) = \{u_1, \ldots, u_{K_{e}}\}$.
To determine which portion of the retrieved users is potentially more beneficial to the recommendation task, we utilize the product of LLM assessment thoughts to generate rankings.
Specifically, we employ a fixed text embedder to obtain dense representations as:
\begin{equation}
	\small
\rep{r}_{t} = \texttt{TxtEmb}(r_{t}), ~\rep{e}^f_{i} = \texttt{TxtEmb}(d(H(u_i))), u_i \in \mathcal{N}^{e}(u_t).
\end{equation}
By computing the cosine similarity between the embedding of the assessment output and the embeddings of the retrieved users' behaviors, we quantify the relative importance of each retrieved user based on their alignment with the beneficial CI characteristics identified in the LLM assessment.
Formally, the relevance ranking can be represented as an ordered set $\mathcal{O}(u_t, v_t)$ (i.e., the reasoning-aligned ranking) derived from $\mathcal{N}^{e}(u_t)$ as follows:
\begin{equation}
	\small
\label{eq:ranking}
\begin{aligned}
	\mathcal{O}&(u_t, v_t) = \left< u_{\pi(1)}, u_{\pi(2)}, \ldots, u_{\pi(K_{e})} \right>, \\
	\text{where} ~ &\rep{r}_{t} \otimes \rep{e}^f_{\pi(1)} \geq \rep{r}_{t} \otimes \rep{e}^f_{\pi(2)} \geq \ldots \geq \rep{r}_{t} \otimes \rep{e}^f_{\pi(K_{e})},
\end{aligned}
\end{equation}
where $\pi$ is a permutation of the indices in $\mathcal{N}^{e}(u_t)$.

{\textbf{Reranker Fine-Tuning.}}
The rankings $\mathcal{O}(u_t, v_t)$ provides principled signals for assessing the CI utility of retrieved users in LLM recommendation.
\emnlp{Using these signals, we construct SARE to rerank the retrieved similar users based on their estimated collaborative value in enhancing LLM-based recommendations.}
We begin by introducing the basic recommendation prompt, which does not incorporate CI.
Taking the movie recommendation as an example, the prompt $P^b$ is:
\vspace{-1mm}
\mybox{Basic Recommendation Prompt $P^b$}{
The user has highly rated the following movies: \textit{<HistoryList>}. Based on this information, predict whether the user would enjoy the movie titled \textit{<TargetItem>}. Respond with either 'Yes' or 'No'.
}
\noindent
This prompt, filled with the target user and item, along with the interactions of users in $\mathcal{O}(u_t, v_t)$, is encoded using the encoder of a reranker:
\begin{equation}
	\small
\begin{aligned}
	&\rep{p}_t = \texttt{RerankerEnc}(P^b(d(H(u_t)), d(v_t))), \\
	\rep{e}^a_i &= \texttt{RerankerEnc}(d(H(u_i))), u_i \in \mathcal{O}(u_t, v_t).
\end{aligned}
\end{equation}
To fine-tune the reranker, we adopt the Top-$k$ shifted by $N$ sampling method~\cite{moreira2024nv}. Specifically, the top-ranked user $u_{\pi(1)}$ from $\mathcal{O}(u_t, v_t)$ is treated as the positive sample, while negative samples are selected as follows:
\begin{equation}
	\small
\label{eq:rare_neg}
	\overline{\mathcal{O}}(u_t, v_t) = \texttt{Sample}_{K_{1}}(\mathcal{O}{(u_t, v_t)}[K_2:]),
\end{equation}
where \texttt{Sample} represents the random sampling of $K_1$ users ranked after $K_2$ in $\mathcal{O}(u_t, v_t)$.
Using the positive and negative samples and following ~\citet{moreira2024enhancing}, we fine-tune the vanilla reranker into SARE with InfoNCE~\cite{oord2018representation}:
\begin{equation}
\label{eq:rare}
\small
	\mathcal{L}_{SARE}
	= - \log \frac{g({ \rep{p}_{t}, \rep{e}^{a}_{\pi(1)}})}{g({ \rep{p}_{t}, \rep{e}^{a}_{\pi(1)}}) + \sum_{u_i \in \overline{\mathcal{O}}(u_t, v_t)} g({ \rep{p}_{t}, \rep{e}^{a}_i}) }.
\end{equation}
The resulting SARE can rerank CAR-retrieved users based on their collaborative value, providing more beneficial CI to enhance the recommendation.
\subsection{CI-Augmented Recommendation}
\label{sec:recommendation}
With CAR and SARE prepared, the original interaction corpus can be effectively leveraged to augment LLM-based recommendations.
Specifically, the retrieved similar user behaviors are integrated with natural language as CI through the following CI-augmented recommendation prompt $P^s$:
\mybox{CI-Augmented Recommendation Prompt $P^s$}{
The user has highly rated the following movies: \textit{<HistoryList>}. Other users with similar preferences have given high ratings to the movies: \textit{<SimilarBehaviorList\_1>, ... <SimilarBehaviorList\_$K_{s}$>}. Based on this information, predict whether the user would enjoy the movie titled \textit{<TargetItem>}. Respond with either 'Yes' or 'No'.
}

{\textbf{CI-Augmented Recommendation Tuning.}}
For each target user, the final users are retrieved following:
 \textbf{Step 1}: Retrieve $K_{e}$ similar users from the interaction corpus using CAR;
 \textbf{Step 2}: Rerank the retrieved users from Step 1 using RARE and select the top $K_s$ users, forming the set $\mathcal{S}(u_t)$.
The similar users in $\mathcal{S}(u_t)$, along with the target user, are then represented in natural language and incorporated into the CI-augmented recommendation prompt $P^s$.
To efficiently adapt LLMs for specific recommendation tasks, we introduce a lightweight LoRA~\cite{hu2021lora} component, where only the parameters of LoRA are trainable while the extensive parameters of the LLM remain frozen. 
The LoRA training objective is defined as:
\begin{equation}
\small
\label{eq:lora}
\begin{aligned}
	&\mathcal{C}(u_t) = \{d(H(u_i))|u_i \in \mathcal{S}(u_t)\}, \\
	\min_{\Theta^\prime}&{\ell{\left(\texttt{LLM}_{\Theta+\Theta^\prime}\left(P^s(d(H(u_t)), d(v_t), \mathcal{C}(u_t))\right), y_t\right)}},
\end{aligned}
\end{equation}
where $\Theta$ represents the parameters of the pre-trained LLM, $\Theta^\prime$ represents the parameters of the LoRA component, $\ell$ is the fine-tuning loss and $y_t$ is the interaction label for the target item $v_t$.

{\textbf{CI-Augmented LLM-Based Recommendation.}}
After training, the LLM equipped with the LoRA module takes the CI-augmented recommendation prompt as input to generate predictions. The recommendation process is as follows:
\begin{equation}
	\small
	\hat{y}_t = \texttt{LLM}_{\Theta+\Theta^\prime}\left(P^s(d(H(u_t)), d(v_t), \mathcal{C}(u_t))\right),
\end{equation}
where $\hat{y}_t$ represents the predicted likelihood of the target user $u_t$ liking the item $v_t$.

\section{Experiments}
In this section, we conduct experiments to answer the following research questions:

\noindent\textbf{RQ1}: Can natural language CI enhance the LLM-based recommendation? How does \model perform compared to the state-of-the-art methods?

\noindent\textbf{RQ2}: Can CAR effectively retrieve useful user behaviors? How does the reranking by SARE contribute to the recommendation performance?

\noindent\textbf{RQ3}: How do the number of retrieved users in CAR and the final set of similar users after reranking by SARE impact performance?

\noindent\textbf{RQ4}: Is \model effective with different CRMs?

\noindent\textbf{RQ5}: Is natural language-form CI effectively utilized by LLMs during recommendation?

\begin{table}
  \centering
  \renewcommand\arraystretch{1}
   \tabcolsep 0.05in
    \scalebox{0.8}{\begin{tabular}{c|cc|ccc}
    \toprule
    \multirow{2}[0]{*}{Dataset} & \multicolumn{1}{c}{\multirow{2}[0]{*}{\# Users}} & \multicolumn{1}{c|}{\multirow{2}[0]{*}{\# Items }} & \multicolumn{3}{c}{\# Interactions} \\
         &      &      & \multicolumn{1}{c}{Train} & \multicolumn{1}{c}{Validation} & \multicolumn{1}{c}{Test} \\
         \midrule
    \textbf{\ml} & 839  & 3,256 & 33,891 & 10,401 & 7,331 \\
    \textbf{\games} & 35,812 & 11,987 & 261,144 & 14,423 & 14,424 \\
    \bottomrule
    \end{tabular}}%
    \vspace{-2mm}
    \caption{Data statistics.}
    \vspace{-6mm}
  \label{tab:dataset}%
\end{table}%

\begin{table*}[ht]
  \centering
  \renewcommand\arraystretch{.95}
   \tabcolsep 0.15in
    \scalebox{0.8}{\begin{tabular}{c|c|ccc|ccc}
    \toprule
      \multirow{2}[0]{*}{Categories} & \multirow{2}[0]{*}{Methods} & \multicolumn{3}{c|}{\textbf{\ml}} & \multicolumn{3}{c}{\textbf{\games}} \\
         &      & AUC  & UAUC & Rel. Imp. & AUC  & UAUC & Rel. Imp. \\
         \midrule
    \multirow{6}[0]{*}{\makecell[c]{Conventional\\models}} & MF   & 0.6063 & 0.5810 & 22.61\% & 0.5311 & 0.5263 & 25.65\% \\
         & LightGCN & 0.6114 & 0.6390 & 16.42\% & 0.5480 & 0.5019 & 26.55\% \\
         & DIN  & 0.6360 & 0.5931 & 18.44\% & 0.6189 & 0.5550 & 13.18\% \\
         & GRU4Rec & 0.6958 & 0.6703 & 6.56\% & 0.6134 & 0.5679 & 12.47\% \\
         & CORE & 0.5901 & 0.5599 & 26.58\% & 0.5927 & 0.5605 & 15.21\% \\
         & SASRec & 0.7000 & 0.6794 & 5.53\% & 0.6111 & 0.5800 & 11.54\% \\
         \midrule
 \multirow{6}[0]{*}{\makecell[c]{LLM-based\\models}} 
 & ICL  &  0.5262    & 0.4767 & 45.15\% & 0.5299 & 0.4886 & 28.42\% \\
 & CoRAL & 0.5412 & 0.5014 & 39.62\% & 0.4910 & 0.4872 & 35.82\% \\
         & TALLRec & 0.7242 & 0.6721 & 4.25\% & 0.6763 & \underline{0.6096} & 3.32\% \\
         & LLaRA & 0.7339 & 0.6967 & 1.75\% & 0.6346 & 0.6035 & 7.31\% \\
         & CoLLM & 0.7333 & 0.6959 & 1.85\% & \underline{0.6797} & 0.5854 & 5.02\% \\
         & BinLLM & \textbf{0.7423} & \underline{0.7044} & 0.62\% & 0.6756 & 0.5961 & 4.47\% \\
         \midrule
         Ours & \model & \underline{0.7367} & \textbf{0.7190} & - & \textbf{0.6933} & \textbf{0.6353} & - \\
         \bottomrule
    \end{tabular}}%
    \vspace{-1mm}
    \caption{Overall performance comparison. \textit{Rel. Imp.} denotes the relative improvement of \model over the baselines based on the average performance of the two metrics. The best results are in bold, and the second-best are underlined.}
    \vspace{-4mm}
  \label{tab:overall}%
\end{table*}%

\subsection{Experimental Settings}
\noindent{\textbf{Datasets.}}
We conduct experiments on two public datasets: MovieLens-1M~\cite{harper2015movielens} (\textbf{\ml}) and Amazon-\textbf{Games}~\cite{ni2019justifying}.
The statistics are shown in \cref{tab:dataset}.
The dataset processing details are in \cref{app:dataset}.

{\textbf{Baselines.}}
We compare \model with two categories of baselines: 1) conventional recommendation models; 2) LLM-based recommendation models.
Specifically, \textbf{1)} conventional recommendation models include Matrix Factorization (\textbf{MF}~\cite{rendle2012bpr}), \textbf{LightGCN}~\cite{he2020lightgcn}, \textbf{DIN}~\cite{zhou2019deep}, \textbf{GRU4Rec}~\cite{hidasi2015session}, \textbf{CORE}~\cite{hou2022core} and \textbf{SASRec}~\cite{kang2018self}.
\textbf{2)} LLM-based recommendation models include \textbf{ICL}~\cite{dai2023uncovering}, \textbf{CoRAL}~\cite{wu2024coral}, \textbf{TALLRec}~\cite{bao2023tallrec}, \textbf{LLaRA}~\cite{liao2024llara}, \textbf{CoLLM}~\cite{zhang2023collm} and \textbf{BinLLM}~\cite{zhang2024text}.
The baseline descriptions are in \cref{app:baseline}.

{\textbf{Evaluation Protocols.}}
Following \cite{zhang2023collm, zhang2024text}, we evaluate the accuracy of these methods in predicting the interaction likelihood of the target item using two commonly employed metrics: AUC and UAUC.
Specifically, AUC measures the overall prediction accuracy, while UAUC computes the AUC score for each user and averages across all users, reflecting the user-level ranking quality.
Higher values indicate better performance.

The implementation details are in \cref{app:implementation}.

\subsection{Overall Performance~(RQ1)}
We conducted experiments to evaluate our framework in enhancing LLM-based recommendation. The overall performance of \model compared to baseline models is presented in \cref{tab:overall}. Several key insights emerged from the results:
\begin{itemize}[leftmargin=6mm]
    \item 
    Our proposed \model achieves the best performance on both metrics for the \games dataset. It also outperforms all models on UAUC for the \ml dataset. Furthermore, \model demonstrates promising relative improvements on averaged metrics. These results substantiate the effectiveness of \model in augmenting LLM-based recommendations through user behavior retrieval and natural language representation.
    \item \model outperforms LLM-based methods that incorporate CI in most cases. Specifically, LLaRA and CoLLM use soft tokens projected from CRM collaborative embeddings as CI. However, they exhibit a significant gap between the collaborative and textual modalities. On the other hand, BinLLM encodes CI in a text-like form (\ie binary strings), but this approach fails to align with LLMs' semantic understanding since such strings rarely reflect the collaborative features found in the original LLM pretraining corpus. In contrast, \model provides CI in a format that aligns well with LLM semantics, thereby enhancing the LLM’s comprehension of the information and improving recommendation accuracy.
    \item LLM-based models generally outperform conventional methods. This is due to LLM-based models benefit from extensive pre-trained knowledge. LLMs can better understand the semantics behind user behaviors, leading to superior recommendation outcomes. However, the ICL and CoRAL, which rely on a general LLM, perform poorly, often underperforming even relative to CRMs. This highlights the limitations of general LLMs in recommendation tasks, which require user behavior context.
\end{itemize}
\subsection{CAR and SARE Contributions~(RQ2)}
\begin{figure}[t]
  \centering
  \begin{subfigure}[t]{0.47\columnwidth}
      \centering
      \includegraphics[width=\columnwidth]{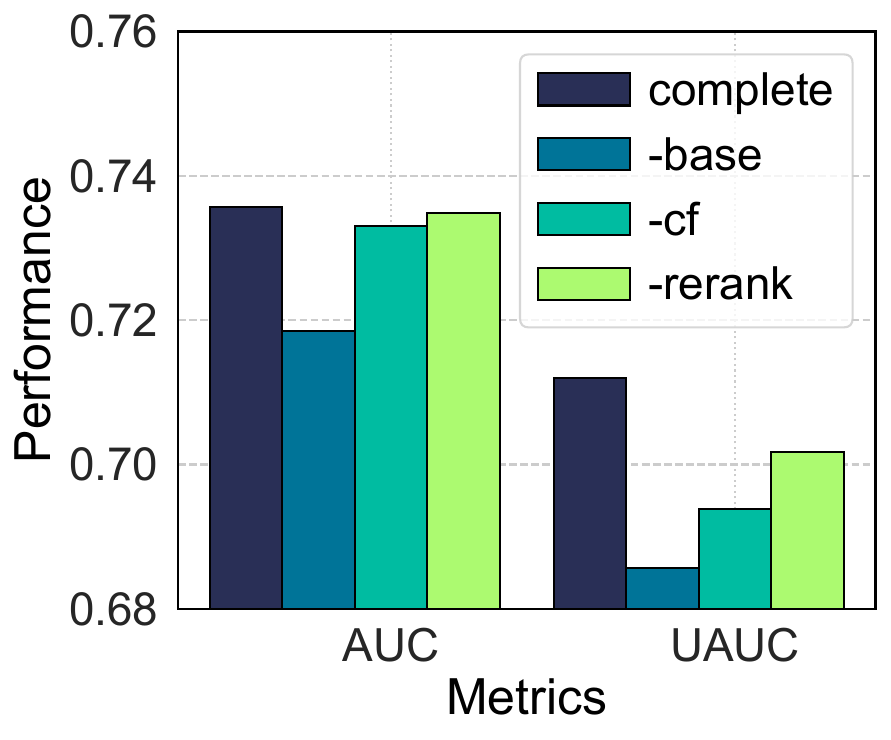}
      \subcaption{ML-1M}
  \end{subfigure}
  \hspace{-1mm}
  \begin{subfigure}[t]{0.47\columnwidth}
      \centering
      \includegraphics[width=\columnwidth]{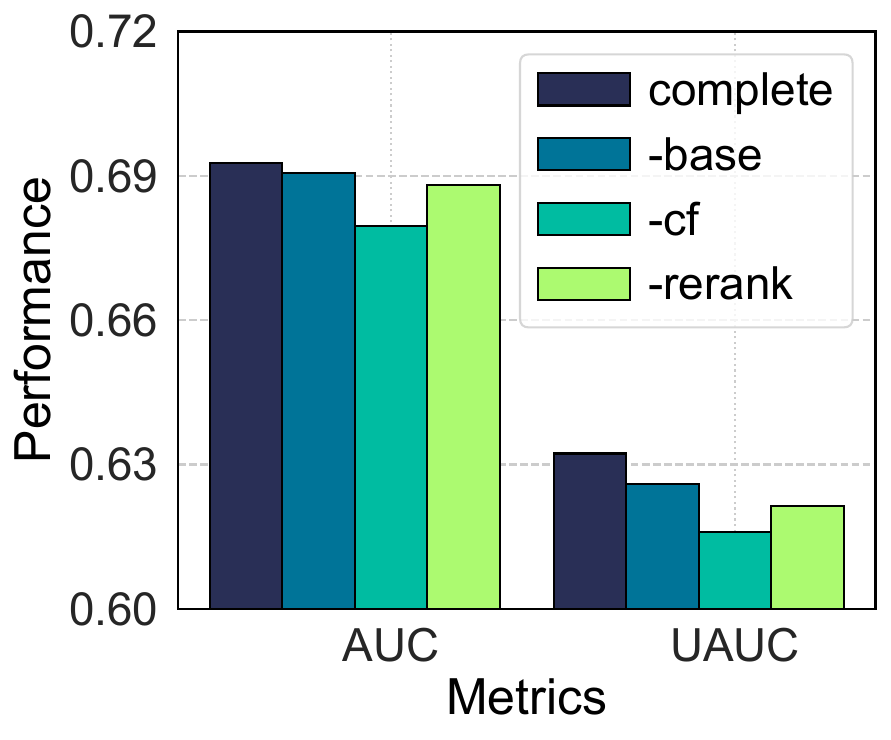}
      \subcaption{Games}
  \end{subfigure}
  \vspace{-3mm}
  \caption{Recommendation performance of different variants of \model.}
  \vspace{-6mm}
  \label{fig:ablation}
\end{figure}

\textbf{Can CAR effectively retrieve useful user behaviors?}
We conducted a comparative analysis with two \model variants:
\textbf{1) \model-base} uses the general retriever to obtain similar users as CI;
\textbf{2) \model-cf} relies solely on a CRM to retrieve similar users based purely on collaborative similarity.
As shown in \cref{fig:ablation}, retrieving similar users based solely on either collaborative similarity (\textit{-cf}) or semantic similarity (\textit{-base}) overlooks important aspects of similarity for LLM-based recommendation.
Pure collaborative similarity fails to capture semantic context, while a general-purpose retriever lacks the domain-specific insights for the task.

\textbf{Can reranking by SARE help the recommendation?} We compare \model with an additional variant: \textbf{\model-rerank} removes SARE for reranking but retains CAR for initial retrieval. 
When the retrieved similar user behaviors from CAR are passed directly to the LLM without reranking, they may include instances that, although similar to the target user, are not beneficial for the LLM's recommendation process. This misalignment can negatively affect performance by introducing irrelevant or less useful information.

Overall, \model outperforms all the variants, confirming the effectiveness of the CAR and SARE components in retrieving relevant user behaviors that enhance LLM-based recommendation.

\subsection{Impact of Retrieved User Number~(RQ3)}
We further investigate how \textbf{1)} the number of retrieved users by CAR (\ie $K_e$) and \textbf{2)} the number of final similar users to act as CI (\ie $K_s$) impact performance. Specifically, we fixed $K_s$ to 2 and varied $K_e$ from 5 to 20, and fixed $K_e$ to 10 while varying $K_s$ from 1 to 4 during recommendation.

\begin{figure}[t]
  \centering
  \begin{subfigure}[t]{0.46\columnwidth}
    \label{fig:sens_car_ml}
      \centering
      \includegraphics[width=\columnwidth]{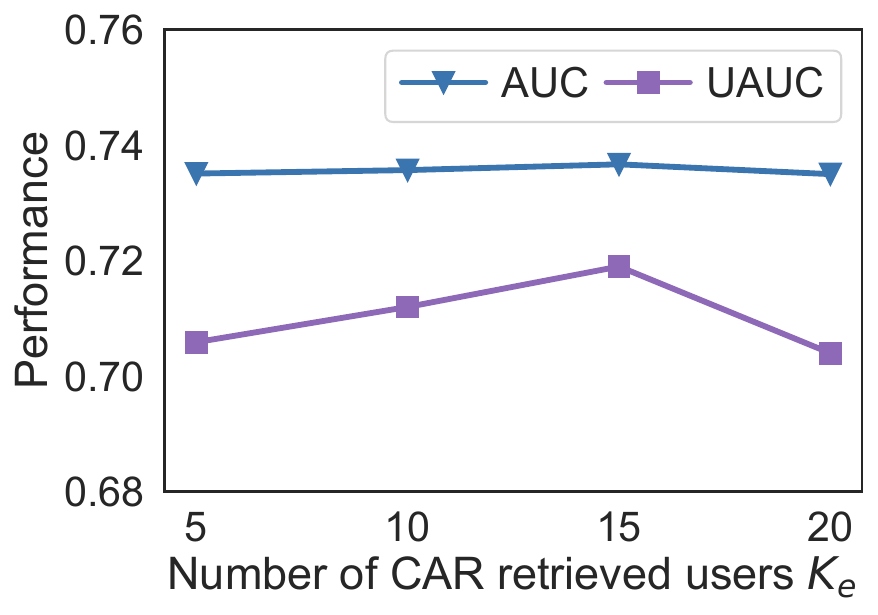}
      \subcaption{ML-1M}
  \end{subfigure}
  \hspace{-1mm}
  \begin{subfigure}[t]{0.46\columnwidth}
    \label{fig:sens_car_games}
      \centering
      \includegraphics[width=\columnwidth]{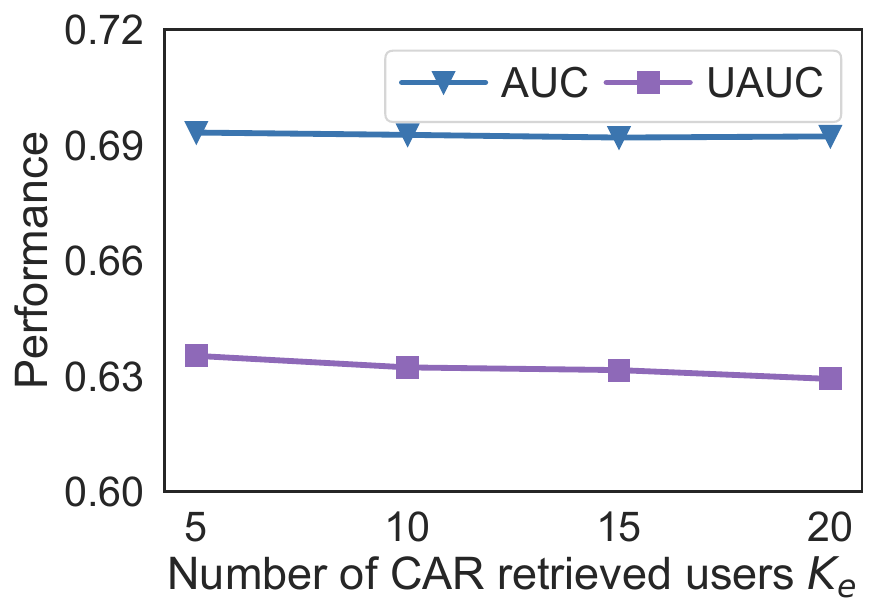}
      \subcaption{Games}
  \end{subfigure}
  \hspace{-1mm}
  \begin{subfigure}[t]{0.46\columnwidth}
    \label{fig:sens_rare_ml}
      \centering
      \includegraphics[width=\columnwidth]{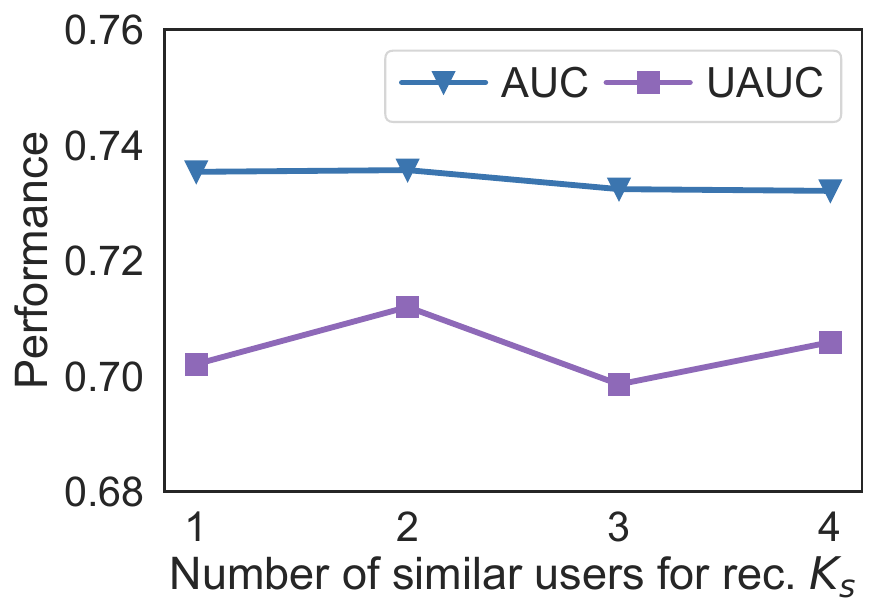}
      \subcaption{ML-1M}
  \end{subfigure}
  \hspace{-1mm}
  \begin{subfigure}[t]{0.46\columnwidth}
    \label{fig:sens_rare_games}
      \centering
      \includegraphics[width=\columnwidth]{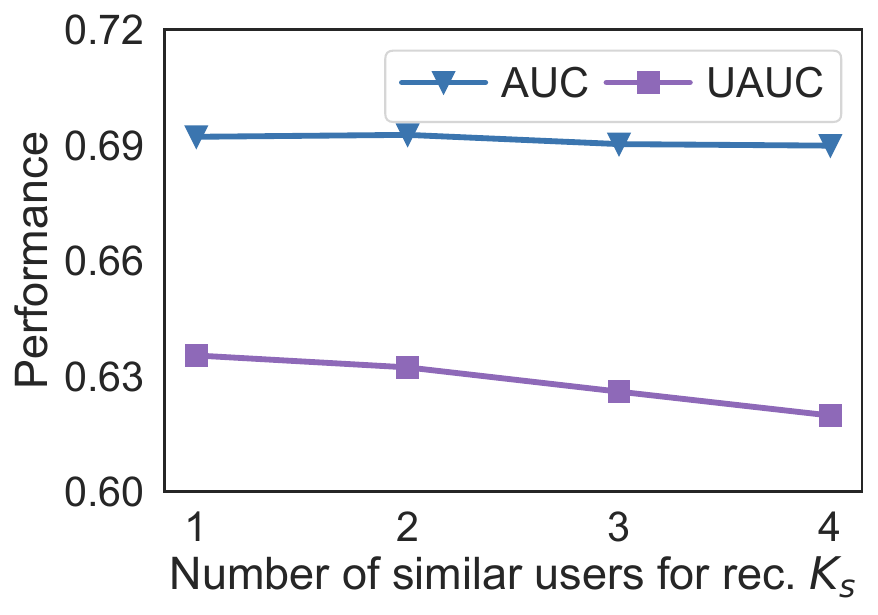}
      \subcaption{Games}
  \end{subfigure}
  \vspace{-3mm}
  \caption{Performance variation with changing the number of retrieved users $K_e$ and final set of users $K_s$.}
  \vspace{-2mm}
  \label{fig:sens}
\end{figure}

Regarding the number of initial retrieved users $K_e$, as shown in \cref{fig:sens_car_ml}, for the \ml dataset, performance initially increases with $K_e$, peaks at $K_e = 15$, and then declines. This pattern suggests that increasing the number of retrieved users allows for richer collaborative knowledge, which benefits recommendations. However, when $K_e$ becomes too large, irrelevant users are included, ultimately detracting from the recommendation performance.
For the \games dataset, shown in \cref{fig:sens_car_games}, performance begins to decline as $K_e$ increases.
This contrast between two datasets suggests that the optimal $K_e$ depends on the dataset's inherent collaborative structure. 

In terms of the number of final similar users for augmenting recommendations, the performance tends to decrease when $K_s$ is large. The performance peaks when $K_s = 2$ for the \ml dataset and $K_s = 1$ for the \games dataset. This decline is likely due to the lost in the middle problem~\cite{liu2024lost}, where overly long contexts cause LLMs to neglect the middle portion.
Thus, providing LLMs with a concise but sufficient amount of CI is crucial for effective recommendation.

\subsection{CRM-Agnostic Analysis~(RQ4)}
We also evaluate the flexibility of \model in integrating with various CRMs to leverage original collaborative knowledge.
Specifically, we select three representative CRMs: SASRec for sequential recommendation, LightGCN for general recommendation, and DIN for CTR prediction.
Additionally, we compare \model with the LLM-based method that is capable of incorporating CI with different CRMs.
Given the architectural similarity between LLaRA and CoLLM, we select LLaRA as it demonstrates superior recommendation performance.

\begin{table}[t]
  \centering
  \renewcommand\arraystretch{1}
   \tabcolsep 0.05in
    \scalebox{0.77}{\begin{tabular}{cccccc}
    \toprule
    \multirow{2}[0]{*}{CRM Backbones} & \multirow{2}[0]{*}{Methods} & \multicolumn{2}{c}{\textbf{\ml}} & \multicolumn{2}{c}{\textbf{\games}} \\
         &      & AUC  & UAUC & AUC  & UAUC \\
         \midrule
    \multirow{2}[0]{*}{SASRec} & LLaRA & 0.7339 & 0.6967 & 0.6346 & 0.6035 \\
         & \model & \textbf{0.7367} & \textbf{0.7190} & \textbf{0.6933} & \textbf{0.6353} \\
         \midrule
    \multirow{2}[0]{*}{LightGCN} & LLaRA & \textbf{0.7288} & \textbf{0.6899} & 0.6141 & 0.5489 \\
         & \model & 0.7206 & 0.6820 & \textbf{0.6940} & \textbf{0.6314} \\
         \midrule
    \multirow{2}[0]{*}{DIN} & LLaRA & 0.7279 & 0.6901 & 0.6789 & 0.6217 \\
         & \model & \textbf{0.7329} & \textbf{0.7009} & \textbf{0.6928} & \textbf{0.6410} \\
         \bottomrule
    \end{tabular}}%
    \vspace{-1mm}
    \caption{Performance comparison between \model and LLaRA with different CRM backbones.}
    \vspace{-6mm}
  \label{tab:crm}%
\end{table}%

As shown in \cref{tab:crm}, \model achieves the best performance on the majority of scenarios across both evaluation metrics. This consistent superiority confirms \model's ability to effectively utilize diverse types of CI from various CRMs, thereby highlighting its flexibility.
Furthermore, we observe that both methods perform better when using SASRec and DIN compared to LightGCN. This disparity likely arises because SASRec and DIN can capture sequential patterns, a capability that LightGCN lacks. Such sequential patterns align naturally with LLM-based methods, which typically represent users based on their historical sequences.

\subsection{Visualization Analysis (RQ5)}
\emnlp{
  To better understand the role of the natural language-form CI introduced in this work, we randomly select an example from the test set of the \ml dataset and visualize the attention weights in the middle layers of the LLM. The input prompt includes the user’s interaction history and retrieved user behaviors, both written in natural language.
  For comparison, we also show the attention patterns for the prompt used in LLaRA, where CI is represented using soft tokens. 
}

\begin{figure}[t]
  \centering
      \includegraphics[width=0.99\columnwidth]{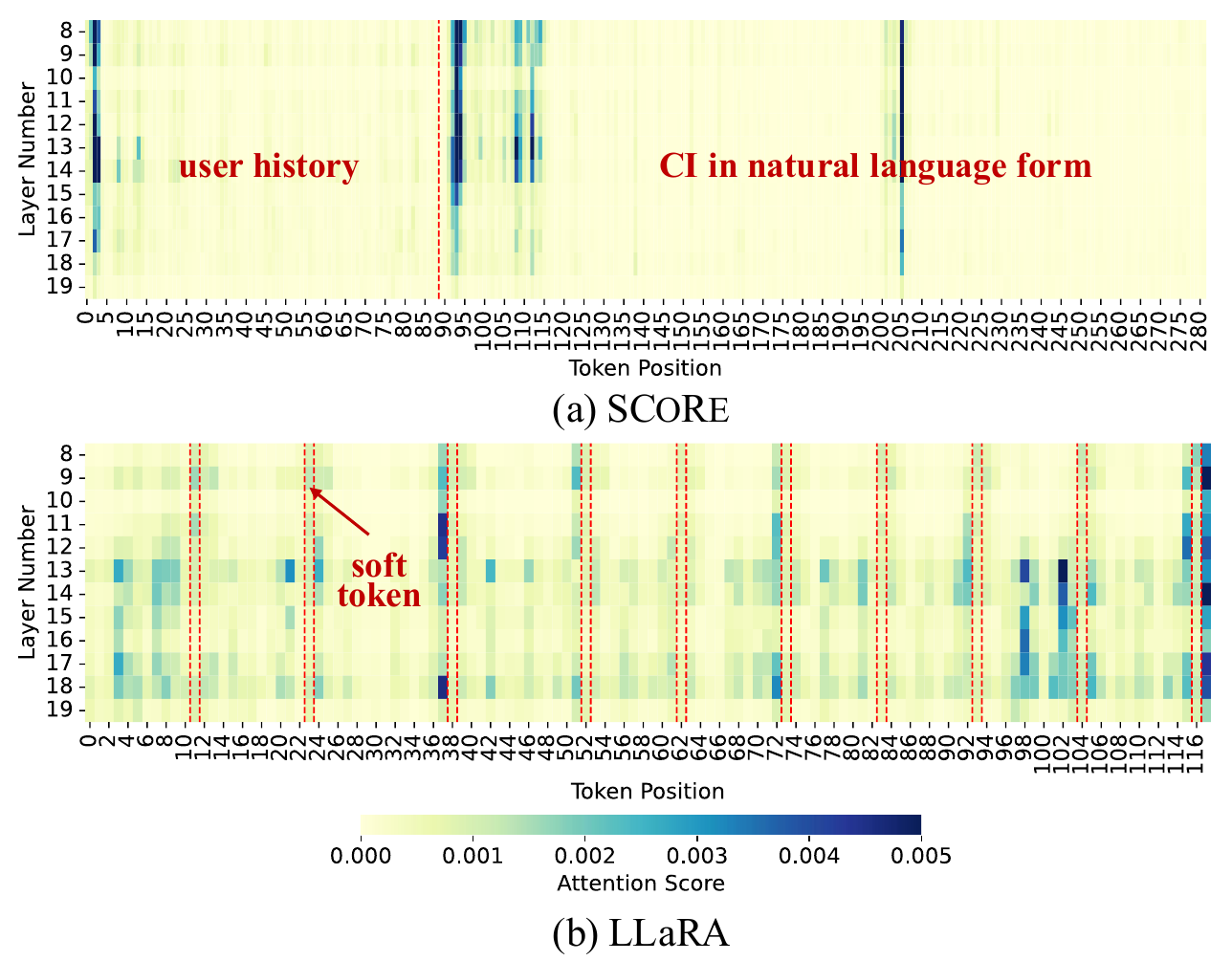}
  \vspace{-7mm}
  \caption{Attention visualization of the input prompt.}
  \vspace{-6mm}
  \label{fig:vis}
\end{figure}

\emnlp{
  As shown in \cref{fig:vis}(a), the model pays varying levels of attention to both the user history and the natural language CI during recommendation. 
  The visible attention weights within the CI suggest that the model is able to use this information effectively. 
  In contrast, \cref{fig:vis}(b) shows that the soft tokens receive little attention, indicating that this format is less compatible with how the LLM-based RSs.
}

\section{Related Works}

\subsection{LLMs for Recommendation}
Recent advancements in LLMs have sparked significant research interest in enhancing RSs.
Several works~\cite{wei2024llmrec,ren2024representation,wang2024can} explored using LLMs as knowledge bases to augment CRMs.
\rev{Instead of separating the recommendation task from the utilization of LLMs,
some studies~\cite{dai2023uncovering,liu2023chatgpt} have directly employed LLMs for the recommendation task, leveraging in-context learning~\cite{dong2022survey} to incorporate demonstration user samples.}
To better adapt general-purpose LLMs for recommendation tasks, numerous studies~\cite{bao2023tallrec,lin2024rella,wang2024can,liao2024llara,zhu2024collaborative} have further explored fine-tuning strategies.
Given the fundamental role of CI in RSs, several works~\cite{liao2024llara,zhu2024collaborative,zhang2023collm,zhang2024text} have incorporated CI during LLM fine-tuning.
\rev{However, these approaches face inherent limitations stemming from modality gaps between CI and the LLM semantic space.
Our work addresses this challenge by representing similar user behaviors as CI in natural language, thereby bridging the semantic gap and minimizing modality divergence.}

\subsection{Retrieval-Augmented Generation}
RAG enhances LLM performance by incorporating relevant information from external knowledge resources~\cite{gao2023retrieval}.
The RAG framework employs either dense~\cite{chen2023dense} or sparse~\cite{ayoub2024case} encoding models to transform user queries into low-dimensional representations.
Given that redundant information can adversely impact generation quality~\cite{liu2024lost}, 
researchers have explored various post-retrieval processing techniques. 
Notable approaches include document chunk reordering~\cite{xu2024list,jia2024bridging,yu2024rankrag} and selective information compression~\cite{jiang2023longllmlingua,xu2023recomp}.
Building upon the RAG framework, our work investigates effectively retrieving similar user behaviors as CI in LLM-based RSs.

\vspace{-1mm}
\section{Conclusion}
\vspace{-1mm}
\emnlp{
In this work, we proposed representing CI in natural language to better align with the semantic space of LLMs, thereby enhancing their recommendation capabilities.
We introduced \model, a novel retrieval framework to identify and integrate relevant user behaviors as supplementary CI.
The collaborative retriever (CAR) captures both collaborative and semantic similarities to retrieve candidate behaviors, while
the self-assessing reranker (SARE) leverages LLM assessments to prioritize those most beneficial for recommendation.
Extensive experiments validate the effectiveness of our approach.}

\section*{Limitations}
Our work has several limitations. First, representing CI in natural language could result in longer prompt length compared to the approaches that use soft tokens or identifiers. In future work, we plan to explore the compression of the CI representations. Second, user behaviors are currently represented in natural language by simply aggregating item metadata. To improve expressiveness, we aim to explore more enriched representations, such as those derived from knowledge graphs or detailed user profiling.


\bibliography{custom}

\appendix
\section*{Appendix}

\section{Dataset Processing}
\label{app:dataset}
For the \ml dataset, we retained interactions from the most recent 20 months, considering ratings greater than 3 as positive feedback.
For the \games dataset, we sampled interactions from 2013 to 2017 and binarized the ratings with a threshold of 4.
To simulate a real-world recommendation scenario and avoid data leakage, we split the datasets into training, validation, and test sets based on temporal order.
Specifically, for the \ml dataset, we used interactions from the first 10 months for training, the following 5 months for validation, and the last 5 months for testing.
For the \games dataset, we used the first 49 months for training and equally split the remaining chronological data into two halves for validation and testing.

\section{Implementation Details}
\label{app:implementation}
When fine-tuning the retriever to craft CAR, we selected mpnet-base\footnote{\href{https://huggingface.co/microsoft/mpnet-base}{https://huggingface.co/microsoft/mpnet-base}} as the foundational retriever and SASRec as the CRM model. The number of similar users, $K_c$, was set to 5, and the batch size was set to 16.
In the LLM-guided alignment, we randomly sampled 10,000 users and used the GLM-4-Plus\footnote{\href{https://bigmodel.cn/}{https://bigmodel.cn/}} API to generate the beneficial CI assessment. The number of retrieved users, $K_{e}$, for participation in the reranking process by CAR was set to 10.
When fine-tuning the reranker to build RARE, we chose bge-reranker-large\footnote{\href{https://huggingface.co/BAAI/bge-reranker-large}{https://huggingface.co/BAAI/bge-reranker-large}} as the foundational reranker. The ranking threshold, $K_1$, was set to 5, and the number of negative samples, $K_{2}$, was set to 3. Moreover, the temperature values in \cref{eq:car} and \cref{eq:rare} were set to 0.1 and 0.02, respectively.
For the recommendation task, we selected the LLM backbone as Llama-3.1-8B-Instruct\footnote{\href{https://huggingface.co/meta-llama/Llama-3.1-8B-Instruct}{https://huggingface.co/meta-llama/Llama-3.1-8B-Instruct}}.
In particular, for retrieval-augmented recommendation tuning, the number of users finally injected into the prompt, $K_{s}$, was set to 2.
Regarding the LoRA module, we set the rank, $\alpha$, and dropout values to 8, 32, and 0.1, respectively. The target injection modules were set to \textit{k\_proj}, \textit{v\_proj}, \textit{q\_proj}, \textit{o\_proj}, \textit{gate\_proj}, \textit{up\_proj}, and \textit{down\_proj}.
During recommendation, we fetched the respective next-token probabilities for the tokens \textit{"Yes"} and \textit{"No"} and normalized them into standard probabilities using the Softmax operation. The normalized probability was considered the final prediction of whether the user would like the target item.
Additionally, we implemented the conventional models using the RecBole~\cite{recbole} library\footnote{\href{https://recbole.io/}{https://recbole.io/}}.
For all the LLM-based models, we used Llama-3.1-8B-Instruct as the backbone, ensuring alignment with \model for fair evaluation.
We conducted the experiments on NVIDIA A40 48GB GPUs.

\section{Baseline Description}
\label{app:baseline}
\textbf{1)}~Conventional recommendation models:
\begin{itemize}
	\item Matrix Factorization (\textbf{MF}~\cite{rendle2012bpr}) factorizes the user-item interaction matrix into latent factors, which represent the users and items in embedding space.
	\item \textbf{LightGCN}~\cite{he2020lightgcn} models user-item interactions as a graph and utilizes a simplified graph convolutional neural network to learn representations.
	\item \textbf{DIN}~\cite{zhou2019deep} employs attention network to learn user interest for Click-Through-Rate~(CTR)  prediction.
	\item \textbf{GRU4Rec}~\cite{hidasi2015session} uses RNNs to model the sequential interactions for the session-based recommendation.
	\item \textbf{CORE}~\cite{hou2022core} takes the linear combination of user sequences and uses a distance measuring method to prevent overfitting.
	\item \textbf{SASRec}~\cite{kang2018self} utilizes a self-attention-based sequential model to better capture long-term semantics.
\end{itemize}

\textbf{2)}~ LLM-based recommendation models:
\begin{itemize}
	\item \textbf{ICL}~\cite{dai2023uncovering} directly uses an untuned LLM to provide recommendations based on user historical interactions.
	\item \textbf{CoRAL}~\cite{wu2024coral} finds the optimal collaborative interactions with a retrieval policy learned through a reinforcement learning framework.
	\item \textbf{TALLRec}~\cite{bao2023tallrec} proposes a two-stage tuning framework including alpaca tuning and rec-tuning to align LLMs with recommendation tasks.
	\item \textbf{LLaRA}~\cite{liao2024llara} proposes a hybrid prompting method that integrates item embeddings from CRMs to better capture behavioral patterns of users.
	\item \textbf{CoLLM}~\cite{zhang2023collm} treats CI as a distinct modality and tunes a mapping module to align it with LLM's token space for recommendation.
    \item \textbf{BinLLM}~\cite{zhang2024text} represents the CI as binary strings generated by deep hashing with graph neural networks.
\end{itemize}

\section{LLM Assessment Example}
\label{app:example}
We randomly sampled a test user from the \ml dataset and presented the user's historical information alongside the assessment generated by the LLM. As shown in \cref{fig:reasoning}, the assessment offers valuable insights into the characteristics of beneficial CI, helping to make more informed decisions about the final recommendation.
\begin{figure*}[t]
    \centering
    \setlength{\fboxsep}{2pt} 
    \fbox{\parbox{0.95\textwidth}{
    {\small\setstretch{1}
        \#\#\# \textit{Input} \\
        The user has watched the following movies: ['Dogma (1999)', 'Clerks (1994)', '12 Angry Men (1957)', 'Talented Mr. Ripley, The (1999)', 'Batman: Mask of the Phantasm (1993)', 'Seven (Se7en) (1995)', 'Fight Club (1999)', 'Annie Hall (1977)', 'Asphalt Jungle, The (1950)', 'Blood Simple (1984)', 'Deer Hunter, The (1978)']. However, based on this list alone, it is not possible to confidently predict whether they would enjoy the movie 'Edward Scissorhands (1990)'. What other genres or movie characteristics related to their preferences, apart from the given history, could help in making a more informed decision? \\\\
        \#\#\# \textit{Response} \\
        To better predict if the user would enjoy "Edward Scissorhands (1990)", consider these additional movie types:

1. \textbf{Fantasy/Dark Fantasy}: "Edward Scissorhands" is a blend of fantasy and dark themes. Movies like "The Crow (1994)" or "Labyrinth (1986)" could indicate a taste for this genre.

2. \textbf{Tim Burton Films}: Since "Edward Scissorhands" is directed by Tim Burton, other films by him, such as " Beetlejuice (1988)" or "The Nightmare Before Christmas (1993)", would be relevant.

3. \textbf{Romantic Dramas with Quirky Elements}: Films like "Eternal Sunshine of the Spotless Mind (2004)" or "Her (2013)" share a blend of romance and unconventional storytelling.

4. \textbf{Coming-of-Age Stories with Unique Protagonists}: Movies like "The Perks of Being a Wallflower (2012)" or "Little Miss Sunshine (2006)" feature atypical main characters and themes of growth.

5. \textbf{Visually Stylish Films}: If the user enjoys visually distinct movies like "Sin City (2005)" or "Pan's Labyrinth (2006)", they might appreciate the aesthetic of "Edward Scissorhands".

These types of movies share thematic, stylistic, or directorial elements with "Edward Scissorhands", providing a more nuanced understanding of the user's potential preferences.
    }}}
    \vspace{-1mm}
    \caption{Example of the user movie viewing history, the target item and the reasoning evidence generated by LLMs. The LLMs provide valuable insights into key characteristics that inform and enhance the recommendation decision.}
    \vspace{-3mm}
    \label{fig:reasoning}
\end{figure*}
\end{document}